\documentclass[aps,prl,twocolumn,superscriptaddress,nofootinbib,10pt]{revtex4-2}

\usepackage{amsmath,amssymb,mathrsfs,bm}
\usepackage{graphicx}
\usepackage{xcolor}
\usepackage[colorlinks=true,linkcolor=blue,citecolor=blue,urlcolor=blue]{hyperref}
\usepackage{orcidlink}

\newcommand{\Lie}{\mathcal{L}}
\newcommand{\CparL}{\mathfrak{C}_{\ell}}
\newcommand{\CparN}{\mathfrak{C}_{n}}
\newcommand{\Ccross}{\mathfrak{C}_{\times}}
\newcommand{\SCross}{\mathfrak{S}_{\times}}
\newcommand{\ThetaW}{\Theta^{\rm W}}

\begin{document}

\title{Local obstructions to trapped-region escape in modified gravity}

\author{Francisco S. N. Lobo}
\email{fslobo@ciencias.ulisboa.pt}
\affiliation{Instituto de Astrof\'isica e Ci\^encias do Espa\c{c}o, Faculdade de Ci\^encias da Universidade de Lisboa, Edif\'icio C8, Campo Grande, P-1749-016 Lisbon, Portugal}
\affiliation{Departamento de F\'isica, Faculdade de Ci\^encias da Universidade de Lisboa, Edif\'icio C8, Campo Grande, P-1749-016 Lisbon, Portugal}

\begin{abstract}
	Energy conditions in modified gravity are commonly imposed on an
	``effective'' stress tensor that mixes physical matter with additional
	gravitational degrees of freedom, obscuring which sector actually controls
	null focusing, horizon multiplicity, and the disappearance of trapped
	regions. We develop a geometry-first alternative in spherical symmetry.
	For a normalized pair of future-directed radial null vectors, two
	longitudinal Ricci projections and one independent transverse Einstein
	projection obey exact area-weighted transport laws. These imply a
	trapping-persistence theorem: after a nondegenerate future-outer marginal
	sphere, a regular ingoing generator cannot exit the trapped region while
	the ingoing longitudinal and transverse convergence channels remain
	nonnegative. In metric $f(R)$ gravity, with $F\equiv df/dR>0$, we then
	construct an entropy-resolved hierarchy based on the local Wald-density
	variable $Fr^2$. In the longitudinal entropy-focusing law, the scalaron
	second derivative cancels exactly, leaving a nonpositive scalar-gradient
	square together with the physical matter projection; in the independent
	transverse entropy-transport law, all scalaron derivative terms cancel.
	Entropy trapping is consequently governed by the physical matter condition
	$T_{nn}\geq0$ and an algebraic mixed threshold rather than by a fictitious
	effective fluid. The resulting hierarchy cleanly separates matter
	positivity, geometric focusing, marginal-surface recovery, and
	gravitational-entropy evolution, and identifies the precise local
	reversals required for regular inner horizons, throats, and
	trapped-to-antitrapped transitions.
\end{abstract}

\maketitle

\emph{Motivation.}---Trapped surfaces are central to black-hole causal structure and to the singularity theorems: the Raychaudhuri equation converts curvature into null focusing, while global assumptions turn that local focusing into statements about geodesic incompleteness~\cite{Raychaudhuri:1953yv,Penrose:1964wq,Hawking:1970zqf,Wald1984}. For any null vector $k^\mu$, the null convergence condition (NCC) is $R_{\mu\nu}k^\mu k^\nu\geq0$, whereas the matter null energy condition (NEC) is $T_{\mu\nu}k^\mu k^\nu\geq0$. In general relativity (GR), Einstein's equation identifies the two. In modified gravity that equivalence fails because additional gravitational degrees of freedom contribute to the same curvature projections as the physical stress tensor. Moving those terms to the right-hand side and calling the result an ``effective'' stress tensor is algebraically convenient, but it can obscure which part of the dynamics is actually matter and which is geometry~\cite{Capozziello:2013vna,Capozziello:2014bqa,Martin-Moruno:2017exc,Kontou:2020bta}.

The distinction becomes decisive when the question is not simply whether a null congruence focuses, but whether a trapped region can \emph{end}. Same-direction focusing is governed by the familiar Raychaudhuri projections. By contrast, recovery of one null expansion while moving in the other future-directed null direction is governed by cross-focusing, the local basis of the outer/inner trapping-horizon classification~\cite{Hayward:1993wb}. This is not a time reversal: both null directions are future directed, and cross-focusing probes a mixed component of the two-dimensional normal geometry. The independence is already visible in Reissner--Nordstr\"om, where a radial Maxwell field saturates both longitudinal radial projections although an inner horizon exists.

This Letter turns that observation into an exact transport hierarchy and then shows that the separation becomes sharper for gravitational entropy. The result addresses a broad class of problems in which one must know \emph{what has to reverse} before a trapped region can disappear: regular black holes, inner horizons, wormhole throats, and black-hole-to-white-hole transitions. Recent kinematical analyses establish that trapped-to-antitrapped transitions require null defocusing~\cite{Borissova:2025msp,Borissova:2026ohj}; the hierarchy below resolves that requirement into distinct longitudinal and transverse mechanisms.

\emph{Three geometric transport channels.}---Write a four-dimensional spherically symmetric geometry as
\begin{equation}
 g_{\mu\nu}=h_{\mu\nu}+r^2\gamma_{\mu\nu},
 \label{eq:metric-prl}
\end{equation}
where $h_{\mu\nu}$ is the Lorentzian metric on the two-dimensional normal space, $\gamma_{\mu\nu}$ is the unit two-sphere metric, and $r$ is the areal radius. The symmetry-sphere area is $\mathcal A=4\pi r^2$. Let $\ell^\mu$ and $n^\mu$ be future-directed radial null vectors normalized by
\begin{equation}
 \ell^\mu\ell_\mu=n^\mu n_\mu=0,
 \qquad \ell^\mu n_\mu=-1.
 \label{eq:nullnorm-prl}
\end{equation}
For any scalar $X$, $\ell(X)\equiv\ell^\mu\nabla_\mu X$ and similarly for $n(X)$; $\Lie_k$ denotes the Lie derivative along $k^\mu$. The null expansions are
\begin{equation}
 \theta_{(\ell)}=\frac{1}{\mathcal A}\Lie_\ell\mathcal A
 =\frac{2}{r}\ell(r),
 \qquad
 \theta_{(n)}=\frac{1}{\mathcal A}\Lie_n\mathcal A
 =\frac{2}{r}n(r).
 \label{eq:expansions-prl}
\end{equation}
A future trapped sphere has both expansions negative. A future marginal sphere of the outgoing family has $\theta_{(\ell)}=0$ and $\theta_{(n)}<0$; it is outer when $\Lie_n\theta_{(\ell)}<0$ and inner when $\Lie_n\theta_{(\ell)}>0$.

For cross-transport along an ingoing generator we choose $n^\mu$ affine, $n^\nu\nabla_\nu n^\mu=0$, and parallel transport $\ell^\mu$ along it, $n^\nu\nabla_\nu\ell^\mu=0$. Spherical radial congruences are shear free and twist free. The cross-focusing equation is then
\begin{equation}
 \Lie_n\theta_{(\ell)}
 =-\theta_{(\ell)}\theta_{(n)}-\frac{1}{r^2}+G_{\ell n},
 \label{eq:cross-prl}
\end{equation}
where $G_{\mu\nu}\equiv R_{\mu\nu}-\tfrac12Rg_{\mu\nu}$, $G_{\ell n}\equiv G_{\mu\nu}\ell^\mu n^\nu$, and $R_{\mu\nu}$ and $R$ are the Ricci tensor and scalar. Define
\begin{equation}
 \CparL\equiv R_{\ell\ell},\qquad
 \CparN\equiv R_{nn},\qquad
 \Ccross\equiv\frac{1}{r^2}-G_{\ell n},
 \label{eq:triplet-prl}
\end{equation}
with $R_{\ell\ell}\equiv R_{\mu\nu}\ell^\mu\ell^\nu$ and $R_{nn}\equiv R_{\mu\nu}n^\mu n^\nu$. The first two are longitudinal; the third is transverse.

The powers of $r$ in the transport laws are fixed by spherical geometry, not chosen ad hoc. Since $r\theta_{(k)}=2k(r)$, multiplication by $r$ cancels the quadratic self-focusing term in the affine Raychaudhuri equation. Since $\Lie_n(r^2)=r^2\theta_{(n)}$, multiplication by $r^2$ cancels the mixed product in Eq.~\eqref{eq:cross-prl}. One obtains the exact identities
\begin{align}
 \Lie_\ell\!\left(r\theta_{(\ell)}\right)&=-r\CparL,
 \label{eq:long-l-prl}\\
 \Lie_n\!\left(r\theta_{(n)}\right)&=-r\CparN,
 \label{eq:long-n-prl}\\
 \Lie_n\!\left(r^2\theta_{(\ell)}\right)&=-r^2\Ccross.
 \label{eq:cross-transport-prl}
\end{align}
Thus $r\theta_{(k)}$ is proportional to the radial velocity, whereas $r^2\theta_{(\ell)}$ is proportional to the outgoing area flux. The three identities are purely geometric: no field equation or matter model has been used.

Let $n^\mu=dx^\mu/d\lambda$ on a finite ingoing affine segment. If two outgoing-family marginal spheres lie on the same regular segment, Eq.~\eqref{eq:cross-transport-prl} gives
\begin{equation}
 \int_{\lambda_a}^{\lambda_b}r^2\Ccross\,d\lambda=0.
 \label{eq:balance-prl}
\end{equation}
Hence a continuous, sign-definite, nontrivial transverse channel cannot connect such a pair.

More strongly, let the generator issue from a nondegenerate future-outer marginal sphere. Here \emph{nondegenerate} means $\Lie_n\theta_{(\ell)}\neq0$ at the marginal sphere, and a \emph{regular segment} means $r>0$ with finite, continuous metric, null-frame, expansion, and curvature data. If subsequently
\begin{equation}
 \CparN\geq0,
 \qquad
 \Ccross\geq0,
 \label{eq:area-persist-prl}
\end{equation}
then $r\theta_{(n)}$ and $r^2\theta_{(\ell)}$ are nonincreasing. Since $\theta_{(n)}<0$ initially and the outer condition makes $\theta_{(\ell)}$ negative immediately inside, both remain negative. No later regular marginal sphere of either radial family can occur before the generator terminates. A regular trapped-region exit therefore requires
\begin{equation}
 \CparN<0\quad\hbox{or}\quad\Ccross<0
 \label{eq:area-reversal-prl}
\end{equation}
somewhere along the relevant segment. This is the first central result: a single NEC-like statement is replaced by an exact identification of the independent geometric channel that must reverse.

\emph{Matter versus geometry in metric $f(R)$ gravity.}---We use units $G=c=\hbar=1$ and consider
\begin{equation}
 S=\frac{1}{16\pi}\int d^4x\sqrt{-g}\,f(R)+S_m,
 \label{eq:action-prl}
\end{equation}
where $S_m$ is the matter action and $T_{\mu\nu}\equiv-2(\sqrt{-g})^{-1}\delta S_m/\delta g^{\mu\nu}$ is the physical matter stress tensor. Let
\begin{equation}
 F\equiv f_R\equiv\frac{df}{dR}>0,
 \label{eq:F-prl}
\end{equation}
so that the effective gravitational coupling is positive. The field equations are
\begin{equation}
 FG_{\mu\nu}=8\pi T_{\mu\nu}+\nabla_\mu\nabla_\nu F
 -g_{\mu\nu}\Box F+\frac12(f-FR)g_{\mu\nu},
 \label{eq:fR-prl}
\end{equation}
where $\Box F\equiv g^{\mu\nu}\nabla_\mu\nabla_\nu F$. For vectors $a^\mu,b^\mu$, write $T_{ab}\equiv T_{\mu\nu}a^\mu b^\nu$ and $\nabla_a\nabla_bF\equiv a^\mu b^\nu\nabla_\mu\nabla_\nu F$.

The area channels become
\begin{align}
 F\CparL&=8\pi T_{\ell\ell}+\nabla_\ell\nabla_\ell F,
 \\[-2pt]
 F\CparN&=8\pi T_{nn}+\nabla_n\nabla_n F,
 \label{eq:fR-long-prl}\\
 F\Ccross&=\frac{F}{r^2}-8\pi T_{\ell n}
 -\nabla_\ell\nabla_nF-\Box F
 \notag\\[-2pt]
 &\hspace{1.9cm}-\frac12(FR-f).
 \label{eq:fR-cross-prl}
\end{align}
Thus the physical matter NEC does not determine the area hierarchy. The additional scalar degree of freedom $F$, the scalaron, can supply geometric defocusing even when matter satisfies the NEC. This is precisely what an effective-fluid rewriting tends to hide.

\emph{Entropy-resolved transport.}---For a stationary horizon in metric $f(R)$ gravity, Wald entropy is~\cite{Wald:1993nt,Iyer:1994ys}
\begin{equation}
 S_{\rm W}=\frac{F\mathcal A}{4}=\pi Fr^2.
 \label{eq:Wald-prl}
\end{equation}
Away from stationarity we use $Fr^2$ as the local Wald-density and conformal-area variable, not as a claim for the most general dynamical entropy. This variable is closely connected with generalized-expansion and dynamical-entropy constructions~\cite{Nielsen:2011zy,Hollands:2024vbe,Kong:2024sqc,Furugori:2025pmn,Jia:2025tgf,Qi:2026imw}. Define
\begin{equation}
 \ThetaW_{(k)}\equiv\frac{1}{Fr^2}\Lie_k(Fr^2)
 =\theta_{(k)}+k(\ln F).
 \label{eq:Theta-prl}
\end{equation}
An entropy-trapped sphere has both entropy expansions negative; an outgoing entropy-marginal sphere has $\ThetaW_{(\ell)}=0$.

Let $k^\mu=dx^\mu/d\lambda$ be affine. Combining the Raychaudhuri equation with Eq.~\eqref{eq:fR-prl} gives
\begin{eqnarray}
 \frac{d\ThetaW_{(k)}}{d\lambda}
 ={}-\frac12\theta_{(k)}^2
 -\left(\frac{1}{F}\frac{dF}{d\lambda}\right)^2
 -\frac{8\pi}{F}T_{kk}.
 \label{eq:entropy-long-prl}
\end{eqnarray}
The scalaron acceleration $d^2F/d\lambda^2$, which enters ordinary area focusing, cancels exactly; only a manifestly nonpositive first-derivative square remains. Therefore $F>0$ and the \emph{physical} matter NEC make $\ThetaW_{(k)}$ nonincreasing. In particular, $Fr^2$ cannot undergo a nondegenerate local minimum along a future affine radial null geodesic unless the physical matter NEC fails. Area and entropy bounces are therefore inequivalent.

The strongest simplification occurs in the independent transverse sector. Define the outgoing entropy flux transported along $n^\mu$,
\begin{equation}
 \mathcal J_{\ell|n}\equiv Fr^2\ThetaW_{(\ell)}
 =\Lie_\ell(Fr^2).
 \label{eq:J-prl}
\end{equation}
Using spherical symmetry and Eq.~\eqref{eq:fR-prl} gives the exact identity
\begin{equation}
 \Lie_n\mathcal J_{\ell|n}
 =-F+r^2\left[8\pi T_{\ell n}+\frac12(FR-f)\right].
 \label{eq:entropy-cross-prl}
\end{equation}
Here \emph{every derivative of the scalaron cancels}. Defining
\begin{equation}
 \SCross\equiv\frac{F}{r^2}-8\pi T_{\ell n}-\frac12(FR-f),
 \label{eq:Scross-def-prl}
\end{equation}
we have $\Lie_n\mathcal J_{\ell|n}=-r^2\SCross$. Thus transverse entropy transport is sourced only by the physical mixed stress, the algebraic $f(R)$ potential, and the intrinsic curvature of the symmetry sphere.

\begin{figure*}[t]
 \centering
 \includegraphics[width=0.92\textwidth]{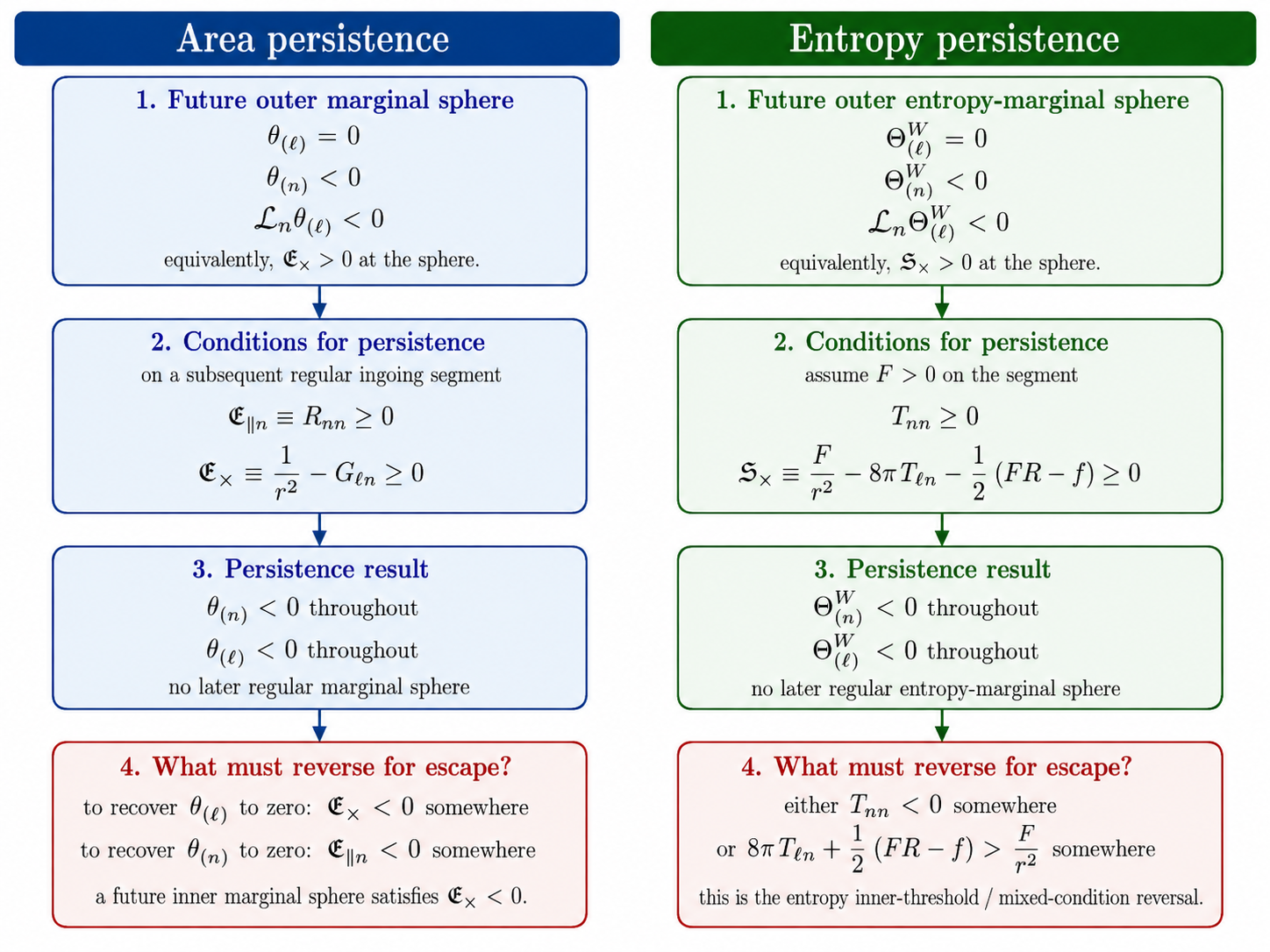}
 \caption{Area versus entropy trapping persistence. Starting from a future-outer marginal sphere, the area hierarchy persists while $\CparN\geq0$ and $\Ccross\geq0$; a regular exit requires the corresponding geometric reversal. For $F>0$, the entropy hierarchy persists instead under the physical matter condition $T_{nn}\geq0$ and the algebraic transverse condition $\SCross\geq0$. The entropy formulation therefore separates genuine matter-NEC violation from a mixed curvature threshold without introducing an effective fluid.}
 \label{fig:persistence-prl}
\end{figure*}

At an entropy-marginal sphere, $\mathcal J_{\ell|n}=0$. A nondegenerate future-outer entropy-marginal sphere has $\SCross>0$, whereas a future-inner one has $\SCross<0$. Hence a future-inner entropy-marginal sphere requires
\begin{equation}
 8\pi T_{\ell n}+\frac12(FR-f)>\frac{F}{r^2}.
 \label{eq:inner-threshold-prl}
\end{equation}
Combining the transverse law with Eq.~\eqref{eq:entropy-long-prl} yields the matter-resolved entropy persistence theorem: from a nondegenerate future-outer entropy-marginal sphere, the conditions
\begin{equation}
 T_{nn}\geq0,
 \qquad
 \SCross\geq0
 \label{eq:entropy-persist-prl}
\end{equation}
keep both entropy expansions negative along the subsequent regular ingoing segment, where regularity now also requires finite $F>0$. A regular entropy-trapped-region exit must therefore involve either a genuine matter-NEC violation, $T_{nn}<0$, or reversal of the algebraic threshold in Eq.~\eqref{eq:inner-threshold-prl}. The area and entropy persistence mechanisms are compared in Fig.~\ref{fig:persistence-prl}.

\emph{Diagnostics and significance.}---The channel independence is not formal. For a radial Maxwell field in asymptotically flat Reissner--Nordstr\"om, with electric charge $Q$,
\begin{equation}
 \CparL=\CparN=0,
 \qquad
 \Ccross=\frac{1}{r^2}-\frac{Q^2}{r^4}.
 \label{eq:RN-C-prl}
\end{equation}
If $r_+>r_-$ are the outer and inner horizon radii, $r_+r_-=Q^2$, then
\begin{equation}
 \Ccross(r_+)>0,
 \qquad
 \Ccross(r_-)<0.
 \label{eq:RN-prl}
\end{equation}
The outer--inner pair is therefore a \emph{pure transverse reversal} despite complete longitudinal saturation. A radial NEC test alone cannot diagnose this horizon multiplicity.

The same mechanism survives with an active scalaron. The exact charged, nonconstant-curvature $f(R)$ solution of Tang, Wang, and Papantonopoulos~\cite{Tang:2019qiy} has, on its positive-coupling inter-horizon domain, $F=1-r/(3m)>1/3$, where $m>0$, and horizons $r_\pm=m\pm\sqrt{m^2-q^2/2}$, with $q$ the charge parameter. Direct null-frame evaluation gives $\CparL=\CparN=0$ while
\begin{equation}
 \Ccross=\frac{1}{2r^2}-\frac{q^2}{4r^4}
 \label{eq:Tang-C-prl}
\end{equation}
reverses sign between $r_+$ and $r_-$. The entropy transverse channel reverses as well, but its source contains no scalaron derivatives.

The same separation clarifies other compact-object scenarios. A nondegenerate null minimum of the areal radius requires negative longitudinal curvature; modified gravity can therefore support an \emph{area} throat with NEC-respecting matter if the extra gravitational sector supplies the defocusing, as exploited in modified-gravity wormhole constructions~\cite{Hochberg:1998ii,Harko:2013yb}. By contrast, Eq.~\eqref{eq:entropy-long-prl} forbids a nondegenerate minimum of $Fr^2$ under the physical matter NEC. Likewise, a trapped-to-antitrapped transition requires longitudinal reversals to change the two expansion signs, while the disappearance and reappearance of marginal boundaries are controlled independently by transverse transport.

The construction is local and quasi-local; it is not a substitute for the global hypotheses of singularity or censorship theorems, nor is an inner marginal surface automatically a stable Cauchy horizon. Classical inner horizons may in fact be destabilized by mass inflation~\cite{Poisson:1990eh}. The strength of the present result is complementary: it gives exact finite-segment obstructions to the recovery of area and entropy expansions and identifies the channel responsible for failure of persistence. These conditions can be tested directly in exact geometries and in double-null numerical evolutions.

\emph{Conclusion.}---The usual question ``does an effective NEC hold?'' is too coarse to determine how a trapped region can disappear in modified gravity. Spherical symmetry reveals three independent transport channels with distinct roles: two govern same-direction null focusing, while the transverse channel governs marginal-surface recovery. Their exact transport laws yield a local trapping-persistence theorem and identify the geometrical reversal required for escape.

For metric $f(R)$ gravity, the entropy variable $Fr^2$ exposes an even sharper separation. The scalaron acceleration cancels from longitudinal entropy focusing, leaving a nonpositive gradient square and the physical matter projection, whereas \emph{all} scalaron derivatives cancel from transverse entropy transport. Entropy persistence is therefore governed by genuine matter positivity and an algebraic mixed curvature threshold, rather than by a fictitious fluid. This is the central physical result: matter positivity, null focusing, horizon multiplicity, and entropy evolution are related but are not interchangeable statements.

The hierarchy provides a direct diagnostic for regular black holes, inner horizons, wormholes, and black-hole-to-white-hole transitions, and it also makes the scope of the present result precise. The \emph{area} transport laws, Eqs.~\eqref{eq:long-l-prl}--\eqref{eq:cross-transport-prl}, and the associated trapping-persistence theorem are purely geometric consequences of spherical symmetry and therefore do not rely on the $f(R)$ field equations. What is specific to metric $f(R)$ gravity is the entropy-resolved reorganization through the local Wald-density variable $Fr^2$, together with the exact cancellation of scalaron derivatives in the transverse entropy channel. In scalar--tensor, Horndeski-type, or more general higher-curvature theories, the natural extension is therefore to identify the appropriate local gravitational-entropy density and determine whether the field equations yield analogous longitudinal and transverse transport laws. The complete derivative cancellation found here should not be assumed a priori; it must be established theory by theory.

Beyond spherical symmetry, the present persistence theorem does not carry over verbatim. The shear-free radial reduction and the unique areal-radius weighting are special to spherical symmetry; for general codimension-two surfaces the null focusing and cross-focusing equations contain shear and normal-bundle contributions, and no single scalar $r$ captures the intrinsic geometry. A natural next step is consequently to formulate the hierarchy directly in terms of the surface area element, the two null expansions, the shears, and the normal-bundle connection/curvature, and then determine which additional terms admit sign control strong enough to imply a generalized persistence statement. These extensions do not modify the local theorem proved here; rather, they identify precisely which structural ingredients must be replaced when one goes beyond metric $f(R)$ gravity or beyond spherical symmetry. Implementing the separate channels in dynamical collapse simulations would provide a direct test of these ideas.\\

\begin{acknowledgments}
FSNL acknowledges funding from the Funda\c{c}\~ao para a Ci\^encia e a Tecnologia (FCT) through national funds under the research grant UID/04434/2025 (DOI 10.54499/UID/04434/2025), and support from the FCT Scientific Employment Stimulus contract with reference CEECINST/00032/2018.
\end{acknowledgments}

\end{document}